\documentclass[
10pt, % Main document font size
a4paper, % Paper type, use 'letterpaper' for US Letter paper
oneside, % One page layout (no page indentation)
headinclude,footinclude, % Extra spacing for the header and footer
BCOR5mm, % Binding correction
]{scrartcl} 
\usepackage{graphicx}
\usepackage[left=3cm,right=3cm,top=3cm,bottom=3cm]{geometry}
\usepackage{float}
\usepackage{url}
\usepackage{authblk}
\usepackage{multicol}
\usepackage{multirow}
\usepackage{hyperref}
\usepackage{xcolor}
\definecolor{yellow}{HTML}{EDC20F}
\definecolor{blue}{HTML}{5B72B7}
\definecolor{turquoise}{HTML}{1DBCC0}
\newcommand{\blue}{\textcolor{blue}}
\newcommand{\yellow}{\textcolor{yellow}}
\newcommand{\turquoise}{\textcolor{turquoise}}

\usepackage{array}
\newcolumntype{L}{>{\arraybackslash}m{3cm}}

\title{Substra: a framework for privacy-preserving, traceable and collaborative Machine Learning}

\author{}
\makeatletter
\renewcommand\@date{{%
  \vspace{-\baselineskip}%
  \large\centering
  \begin{tabular}{@{}c@{}}
    Mathieu Galtier\textsuperscript{1, 2} \\
    \normalsize mathieu.galtier@owkin.com
  \end{tabular}%
  \quad and\quad
  \begin{tabular}{@{}c@{}}
    Camille Marini\textsuperscript{1, 2} \\
    \normalsize camille.marini@owkin.com
  \end{tabular}

  \bigskip

  \textsuperscript{1}Owkin, France\par
  \textsuperscript{2}Substra Foundation, France

  \bigskip

  \today
}}
\makeatother

\begin{document}

\maketitle

%\tableofcontents

\begin{abstract}
Machine learning is promising, but it often needs to process vast amounts of sensitive data which raises concerns about privacy.
In this white-paper, we introduce Substra, a distributed framework for privacy-preserving, traceable and collaborative Machine Learning.
Substra gathers data providers and algorithm designers into a network of nodes that can train models on demand but under advanced permission regimes.
To guarantee data privacy, Substra implements distributed learning: the data never leave their nodes; only algorithms, predictive models and non-sensitive metadata are exchanged on the network.
The computations are orchestrated by a Distributed Ledger Technology which guarantees traceability and authenticity of information without needing to trust a third party.
Although originally developed for Healthcare applications, Substra is not data, algorithm or programming language specific.
It supports many types of computation plans including parallel computation plan commonly used in Federated Learning.
With appropriate guidelines, it can be deployed for numerous Machine Learning use-cases with data or algorithm providers where trust is limited. \newline
\end{abstract}

\textbf{Context} 
Substra is an open source framework which can be found on the Substra github (\href{https://github.com/SubstraFoundation}{https://github.com/SubstraFoundation}). It was originally developped by Owkin (\href{https://www.owkin.com/}{https://www.owkin.com/}), which proposes an enterprise version of Substra for Healthcare. It is now hosted by the nonprofit Substra Foundation (\href{https://www.substra.ai/}{https://www.substra.ai/}).

\section{Introduction}

%\subsection{Motivation}
\textbf{Machine Learning (ML) is a promising field} with many applications; organizations of all sizes are practising it, from individual researchers to the largest companies in the world.
In doing so, they concentrate an extremely large amount of data.
Today, data business is flourishing.
However, these practices raise important ethical questions which ultimately could limit the potential social benefits of ML \cite{bostrom2014ethics, papernot2016towards}.
ML requires large amounts of data to learn from examples efficiently \cite{halevy2009unreasonable}.
In ML more data often leads to better predictive performance.
%However, data can be produced in a distributed way, by many different sources such as users, patients, measuring devices etc.
Usually, different sources, such as users, patients, measuring devices etc, produce data in a decentralized way.
This source distribution makes it difficult to have enough data for training accurate models.
Currently, the standard methodology for ML is to gather data in a central database.

% Talk about the cloud 
% data minimization 

\textbf{However, data is often sensitive}. 
In the case of personal data, which are explicitly related to an individual, privacy is at stake.
Personal data are particulary useful and valuable in the modern economy.
With personal data it is possible to personalize services, which has brought much added value to certain applications.
This can involve significant risks if the data are not used in the interest of the individual.
Not only should personal data be secured from potential attackers, but the organisations collecting data should also be transparent and aligned with user expectations.
In the European Union, the General Data Protection Regulation (GDPR) \cite{GDPR-ref} has imposed consent and control of citizens over their personal data as a fundamental right.
Beyond privacy, data can also be sensitive when it has economic value.
Information is often confidential and data owners want to control who accesses it.
Examples range from classified information and industrial secrets to strategic data which can give an edge in a competitive market.
From the perspective of tooling, privacy-preserving and confidentiality-preserving are very similar and differ mostly in the lack of regulation covering the latter.

Thus, a tradeoff exists between predictive performance improvement versus data privacy and confidentiality.
ML always needs more data, but data tend to be increasingly more protected.
The centralization paradigm where a single actor gathers all data on its infrastructure is reaching its limit.

A relevant way to solve this tradeoff lies in \textbf{distributing computing} and remote execution of predictive tasks.
In this approach, the data themselves never leave their nodes.
In ML, this includes \textbf{Federated Learning}: each dataset is stored on a node in a network, and only the algorithms and predictive models are exchanged between them \cite{mcmahan2016communication, bonawitz2019towards}.
This immediately raises the question of the potential information leaks in a trained model.
The research on ML security and privacy has seen a significant increase in recent years covering topics from model inversion \cite{fredrikson2015model} and membership attacks to model extraction \cite{tramer2016stealing}.
A residual risk is that data controllers still have to trust a central service that orchestrates federated learning, and distributes models and metadata across the network.
Research in Secured Multi Party Computation (SMPC) \cite{goldreich1998secure} has proposed several schemes and tools to solve the problem and some have been recently proposed precisely on the ML context of this whitepaper \cite{hie2018realizing}. The results are promising, but a large computing and communication overhead may slow the growth of this field.

\textbf{Reliability and reproducibility} of ML is also a major challenge to wide social and market adoption.
It is now clear that ML is relevant in many well defined industrial applications, but it is restricted to standardized tasks and still needs to improve in the face of the inherent variability of certain phenomena.
For some sensitive applications, such as Healthcare, one can not tolerate mistakes.
In ML, predictive models are trained from a set of examples and, even with the best technology, models will perform poorly on a new example which may be significantly different from the training dataset.
Building representative datasets is key to creating robust models.
Consequently, it is fundamental to consider the training of predictive models together with sound evaluation. A sound evaluation should always be performed on a representative test dataset of the target population of data.
This evaluation should be entirely traceable and reproducible.
Furthermore, in sensitive situations, evaluation should be done by independent organisations.
Today, there is a lack of collaborative tools which could make the evaluation or certification processes regarding the ML predictions more reliable.

Parallel to the growth of ML and its risk mitigations, the field of \textbf{Distributed Ledger Technologies (DLT)} has recently gained momentum with the rapid rise of blockchain technology from early conception to deployment of mature technology; networks that are both broadly used and indestructible.
Services built on top of blockchains are said to be trustless: one does not need to trust a third party to guarantee integrity and availability of the service.
Today, a large number of users contribute daily to the secure hosting of a distributed and unfalsifiable database, called a ledger, on networks powered by protocols such as Bitcoin \cite{nakamoto2008bitcoin} or Ethereum \cite{wood2014ethereum}.
Ledger networks are often operated through smart-contracts which are simply traceable functions on the state of the ledgher.
This amounts to creating \textit{trustless} services where one does not rely on a third party to provide a reliable service.
In the wake of public blockchains, several private blockchain frameworks have emerged, many of them are hosted by the Hyperledger initiative \cite{Hyperledger-ref, androulaki2018hyperledger}.
The core difference is that private blockchains are deployed within a restricted group of users.
This significantly simplifies the underlying consensus mechanism and, in particular, removes the requirement of large computing power associated with the Proof of Work consensus mechanism, as is required by the Bitcoin network.

In this whitepaper, we describe \textbf{Substra, a traceable and privacy-preserving framework for collaborative ML} which tackles the challenges of robust ML on sensitive data.
It orchestrates the remote execution of ML models over distributed datasets under advanced privacy constraints.
Substra relies on a private DLT to implement distributed learning in a trustless way. It connects several users controlling different datasets, to algorithms providers and independent performance evaluators. In this document, we cover the principles, concepts, usage, architecture, ML orchestration features and risk analysis of the proposed technology.

\section{Principles}
Three core principles drive the development of Substra:
\begin{itemize}
    \item \textbf{Collaboration}.
    In practice, data is often spread among several partners and the algorithmic expertise can belong to yet another institution.
    Substra is rooted in the belief that state of the art ML will be built within networks of partners, in particular when the data is sensitive.
    \item \textbf{Privacy}. Data controllers should never expose their data to obtain a service based on ML.
    Sensitive data should remain private and never be transferred to a third party.
    Favoring remote execution rather than remote access, Substra is decentralized and makes it impossible for anyone but the owner or authorized algorithms to access the data.
    \item \textbf{Traceability}. Complete traceability of all ML operations is essential not only to guarantee privacy of data, but also to provide an untampered history of the training of any predictive model.
    This is necessary to support any reliability claim regarding the performance of a model.
\end{itemize}

\section{Concepts}
Here, we introduce the main concepts underlying Substra. In fact, Substra is a framework to orchestrate computations in different nodes over several \textit{Assets} under the constraint of explicit \textit{permission regimes}.

\subsection{Nodes}
Nodes are standalone computing and storage resources running the Substra code.
They are organised into a network.
It is assumed that independent partner organizations control their respective nodes.
They form a private network, where every node is connected to all others.

In Substra, users are authenticated through the node they belong to (see section \ref{sec: architecture} on architecture).
There are only credentials at the institution level: individual users are not personally identified at the network level.
Thus, throughout the document, we will refer to users, organisations, institution, partners or, nodes indistinctly. 

\subsection{Assets}
Substra registers, stores and organizes computations on four different kinds of \textit{Assets}: Objectives, Datasets, Algorithms and, Models.
These assets can be private or shared depending on their \textit{permission regime}.
\begin{figure}[htbp] 
    \centering
    \includegraphics[width= 0.5\textwidth]{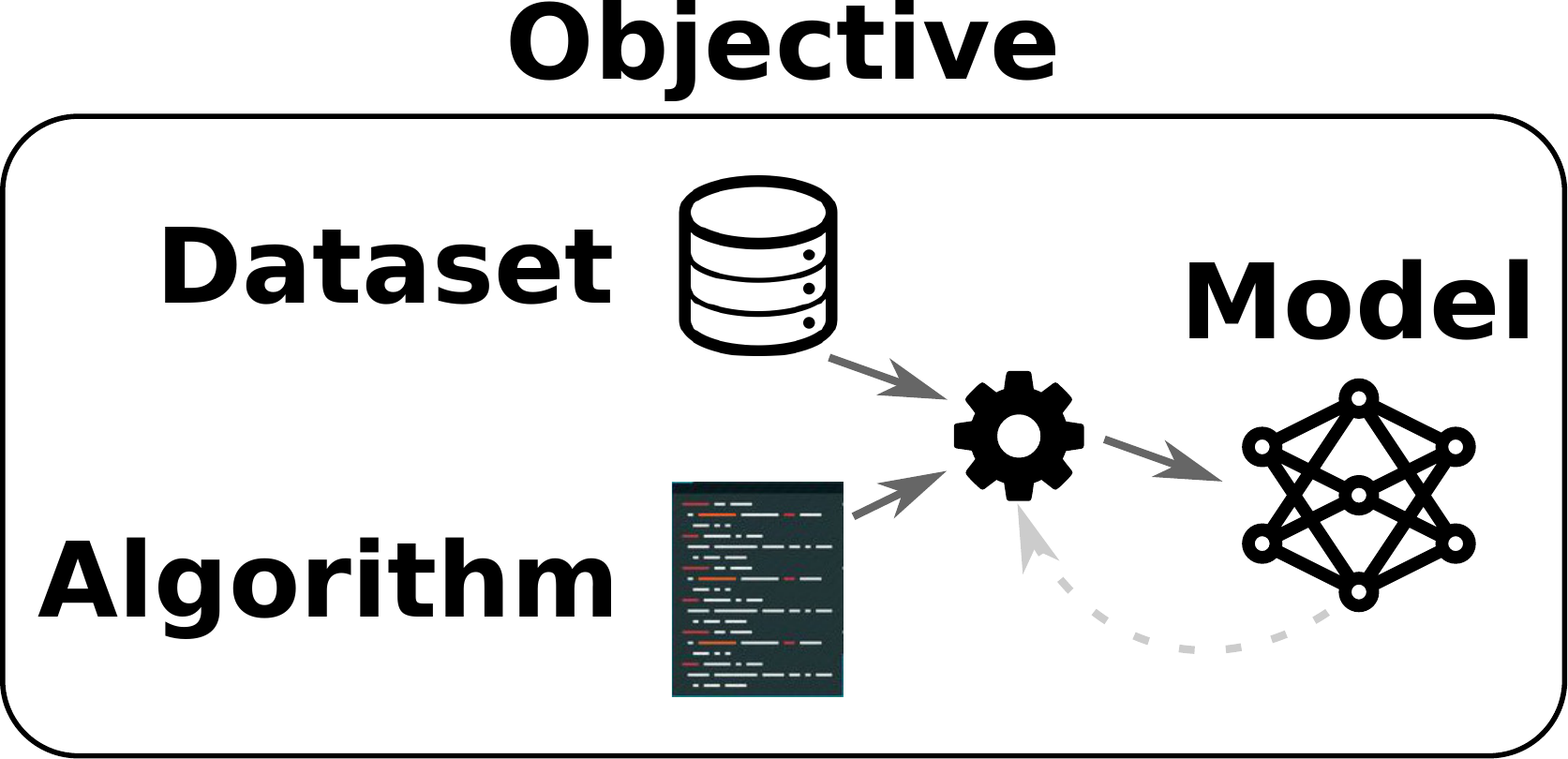}
    \caption{The four types of assets in Substra. }
    \label{fig: assets}
\end{figure}
\begin{itemize}
    \item An \textit{Objective} clearly defines the purpose of the computations.
    It specifies (i) the data format that the \textit{Dataset}, \textit{Algorithm} and \textit{Model} must follow, (ii) the identity of the test data points used to compare and evaluate the models and, (iii) the metric calculation script which is used to quantify the accuracy of a model.
    \item A \textit{Dataset} aggregates numerous data points under a common standard format.
    It includes a single \textit{Opener} script which imports and opens the file using libraries specific to the data type.
    \item An \textit{Algorithm} is a script which specifies the method to train a \textit{Model} on a \textit{Dataset}.
    In particular, it specifies the model type and architecture, the loss function, the optimizer, hyperparameters and, also identifies the parameters that are tuned during training.
    \item A \textit{Model} is a potentially large file containing the parameters of a trained model.
    In the case of a neural network, a model would contain the weights of the connections.
    It is the result of training an \textit{Algorithm} with a given \textit{Dataset}.
    In Substra a \textit{Model} is defined through a training task which is specified by a \textit{traintuple}.
    The \textit{Model} can be evaluated via the defintion of a \textit{testtuple}.
   
\end{itemize}

Substra is agnostic with respect to the \textit{Assets} nature and format.
Substra can be used in any field for any \textbf{supervised Machine Learning problem}.
Consequently, it is up to the users to design and respect consistent interoperability conventions for their specific application.
\textit{Openers} and \textit{Algorithms} have to be manually made compatible for each \textit{Objective}.
The format of each \textit{Asset} provided by users must therefore be documented in detail.

\subsection{Permission regimes}
Each \textit{Asset} in Substra has its own \textit{permission regime}.
A \textit{permission regime} specifies which organizations can process or download a given \textit{Asset}, as described below. 

There are two types of permissions:
\begin{itemize}
    \item The permission to \textbf{process} an \textit{Asset} provides the ability to utilize it in a training or prediction task.
    If permission to process an \textit{Asset} is given to a node, then the latter can request the processing of the \textit{Asset}.
    But the \textit{Asset} never leaves the node of its owner: the processing is done within the owner node.
    For instance, a \textit{Dataset} can be used to train a model by any organization having the process permission.
    \item The permission to \textbf{download} an \textit{Asset} provides the ability to retrieve and access the \textit{Asset}.
    If permission to download an \textit{Asset} is given to a node, then the \textit{Asset} will be shared between this node and the owner node.
    Of course, the point of Substra is that no \textit{Dataset} is ever given a download right (excluding samples of anonymized data points for prototyping). 
\end{itemize}
In Substra, having the download right over an asset implies having the process right.

Beyond the whitelist of organizations having process and/or download rights, an \textit{Asset} can also be made available for processing only for specific purposes.
To do so the \textit{Asset} owner must provide a whitelist of \textit{Objectives} for which the \textit{Asset} can be used.

\textit{Models} that are created by Substra inherit their \textit{permission regime} from the \textit{Assets} that were used during its creation.
By default, the process and download whitelists of the new model are the intersection of the whitelists of the \textit{Dataset}, \textit{Algorithm} and initial \textit{Models} used for training.
For now, at least one organization must be given the download right so as to be able to store the \textit{Asset}.
This choice is made explicitly in the \textit{traintuple} specifying the \textit{Model} creation.

In Substra, the permission regimes are enforced a priori: a \textit{traintuple} can only be created if it respects the \textit{permission regime} of each \textit{Asset} involved.
In other words, computations can only be triggered by an organization having at least the process right over all assets referenced in the \textit{traintuple}.
In fact, the permissions are implemented in trustless smart-contracts which filter the addition of \textit{traintuples} in the ledger.

\subsection{Computations}
At its core, Substra is a tool to orchestrate the execution of training tasks.
These training tasks turn a triplet of \textit{Dataset}, \textit{Algorithm}, and \textit{Model} into an updated \textit{Model} (see figure \ref{fig: assets}).
The goal is to fit the model to a new set of data in order to increase performance on similar data points.
The specification of a training task is entirely contained in a \textit{Traintuple}, which gathers the relevant information about the necessary \textit{Assets} and all technical variables to unequivocally describe a training task.

\textit{Traintuples} have a counterpart for the test of a \textit{Model} on a separate \textit{Dataset}: \textit{Testtuples}. They correspond to the specification of evaluation tasks of \textit{Models} resulting from  \textit{Traintuples}.

One can form a chain of training (and evaluation) tasks, where a model is sequentially updated with various \textit{Datasets} and/or \textit{Algorithms}.
We call such chains of tasks \textit{Compute Plans}.
They can also form more complicated structure with parallelism and pooling steps involved as will be detailed in section \ref{compute plans}.

\section{Usage}
Here, we describe how Substra can be used.
First, the canonical use cases are detailed; second, the unitary operations allowed are listed; third, the interfaces to interact with Substra are described.

\subsection{Use-cases}
Three canonical use-cases for Substra are detailed: the \textit{data / algorithm collaboration}, the \textit{data consortium}, and the \textit{Training / evaluation collaboration}.
The first two use cases rely on the fact that efficient predictive models require lots of data to be trained.
Thus, collaborations can be set up between several actors to increase the amount of data from which a model is trained.

Note that these use cases are compatible. Indeed, all the real world applications of Substra we are currently aware of borrow from the use-cases below.

\subsubsection{Data / algorithm collaboration}
When data controllers and model engineers belong to different organizations, effective collaboration can be a real challenge as data controllers may not be willing to transfer their sensitive data to potentially untrusted model engineers.
Typically, data controllers host large amounts of sensitive data which can only be processed under strong confidentiality constraints.
They have an incentive to limit the number of copies of their data, and are reluctant to provide access to the data itself.
Model engineers design algorithms which often need large amounts of quality data to be used to build predictive models.
Model engineers are always looking for more data to train their models and build new and better services.
Thus, the collaboration would be beneficial to both but transfering the data is often to risky for the data controller.
In this setup, Substra fosters collaboration by addressing and removing the need of data transfer.

For instance, in the precise use-case for which Substra was originally designed hospitals are the data controllers. Through their typical operations they collect countless sensitive and private data.
They want to valorize this data in order to provide better patient-care or to foster medical research.
Yet, they cannot share the data without constraints.
In this use-case, model engineers are either academic researchers or companies specialized in medical AI which want to create and/or sell ML based services.
These predictive services can be provided directly to the hospitals, or to some third party.

\subsubsection{Data consortium}
When competing entities separately collect very similar data, they may be interested in mutually improving the efficiency of their predictive model provided their data remains private.
These organizations may want to collectiveley train an algorithm across their datasets and share the resulting aggregate model.
Such an approach could improve the efficiency of the entire sector without favoring one actor over the other.

For instance, Substra can be used between several pharmaceutical companies which have almost identical processes to discover new drugs and have gathered very similar data over the years.
Crucially, these companies are competitors and want to protect their data from each other.
Yet, using Substra, they can collectively train a common predictive model without revealing their data.
Thus, Substra helps them improve their ability to discover new drugs.

\subsubsection{Training / evaluation collaboration}
The practice of training predictive models is becoming widespread, but rigorous evaluation of performance is not always occuring. In most fields, the rise of ML is conditioned by the concrete proof that the ML models can generalize robustly and can be applied consistently on new data points. To measure the capacity of a ML model to generalize, a simple but efficient way is to evaluate it on a representative dataset it has never seen before: this is a test dataset.

Thus independent evaluators could gather representative, well-curated and non-biased datasets for testing the \textit{Models} a posteriori. The Evaluators would design an \textit{Objective} with a test dataset and open of leaderboard for the predictive models.

For instance in healthcare, one could imagine that regulatory bodies team up with strategic hospitals to register test cohorts so that all models in Substra can be benchmarked independently. For a given pathology, there could be test datasets "certified" by independent organisations to help evaluate \textit{Models}. Then startups could design \textit{Models} which are evaluated on independent data, leading to increased reproducibility.

\subsection{Operations}
Substra makes it possible for a user to
\begin{itemize}
    \item Create an \textit{Asset} such as a \textit{Dataset}, \textit{Algorithm}, or \textit{Objective}, via direct upload or regsitration from file.
    \item Change the \textit{permission regime} of an \textit{Asset}.
    \item Train a \textit{Model} from available \textit{Assets} by creating a \textit{Traintuple}.
    \item Evaluate the performance of a \textit{Model} on the test data of an \textit{Objective}, or using a cross-validation approach on a \textit{Dataset}, by creating a \textit{Testtuple}.
    \item View and compare the performance of all \textit{Models} in the form of a leaderboard (i.e. a list of models ordered by performance).
    \item Request a prediction from a \textit{Model} on a new data point.
\end{itemize}

\subsection{Interfaces}
In order to perform these operations Substra comes with 3 types of interfaces: a web interface, a Command Line Interface (CLI), and a Python Software Development Kit (SDK). 
\begin{itemize}
    \item The \textbf{frontend} aims at providing traceability of all operations on Substra \textit{Assets}. It can also be used to choose the desired \textit{permission regimes} on \textit{Assets}.
    As shown in figure \ref{fig: frontend}, the Substra frontend displays lists of all \textit{Assets} in specific tabs.
    A search bar can be used to filter \textit{Assets}. 
    \begin{figure}[htbp] 
        \centering
        \includegraphics[width= 0.8\textwidth]{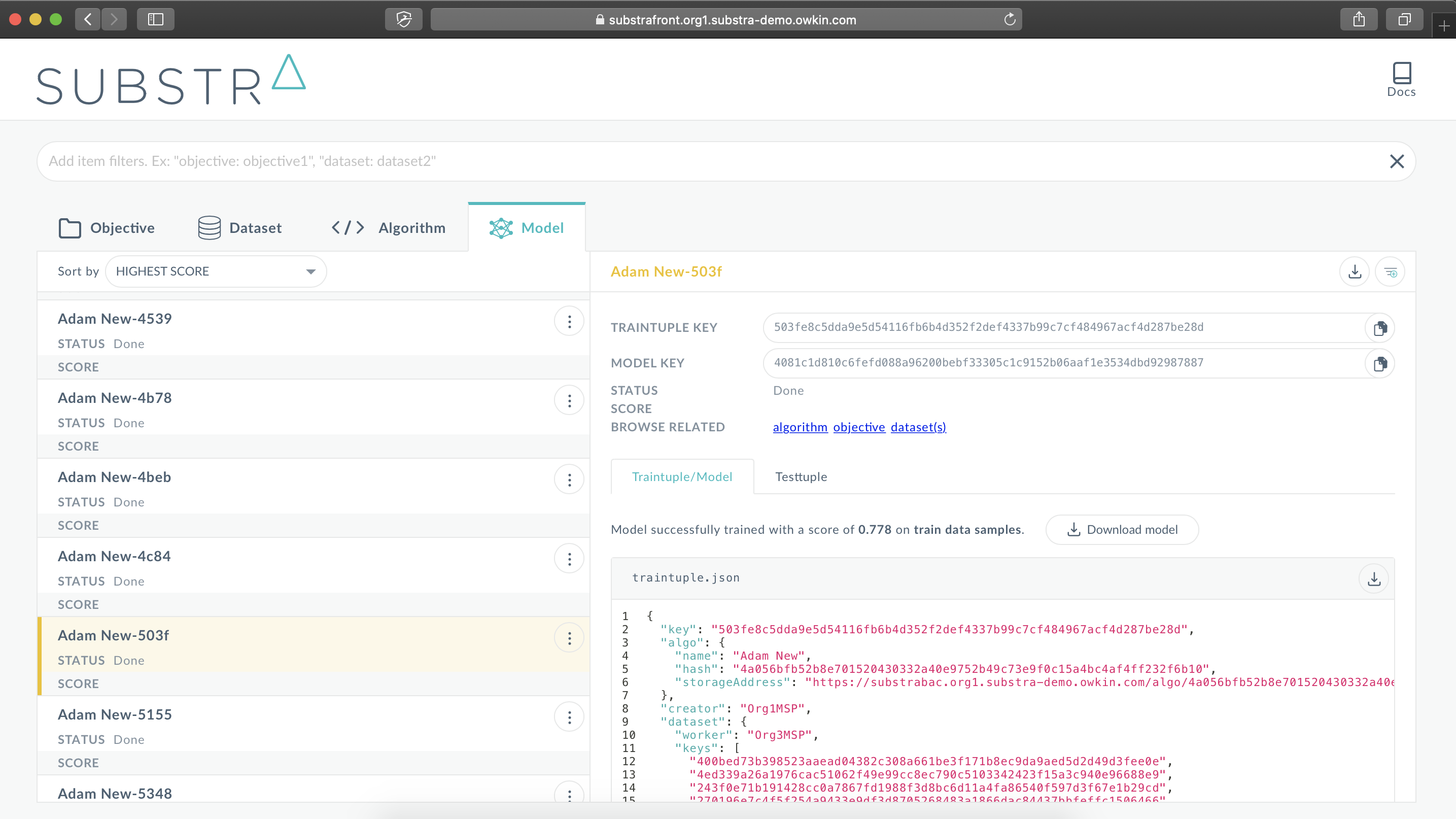}
        \caption{Screenshot of the Substra frontend. The \textit{Model} page of the frontend shown here displays a leaderboard for a given \textit{Objective}.}
        \label{fig: frontend}
    \end{figure}

    \item The \textbf{CLI} makes it possible to add \textit{Assets} to Substra and also to list registered \textit{Assets}.  Figure \ref{fig: cli} shows the commands which can be executed with the CLI. Importantly, each \textit{Asset} must be formatted in a proper way before being pushed to the platform. 
         \begin{figure}[!h] 
       		 \centering
       			 \includegraphics[width= 0.6\textwidth]{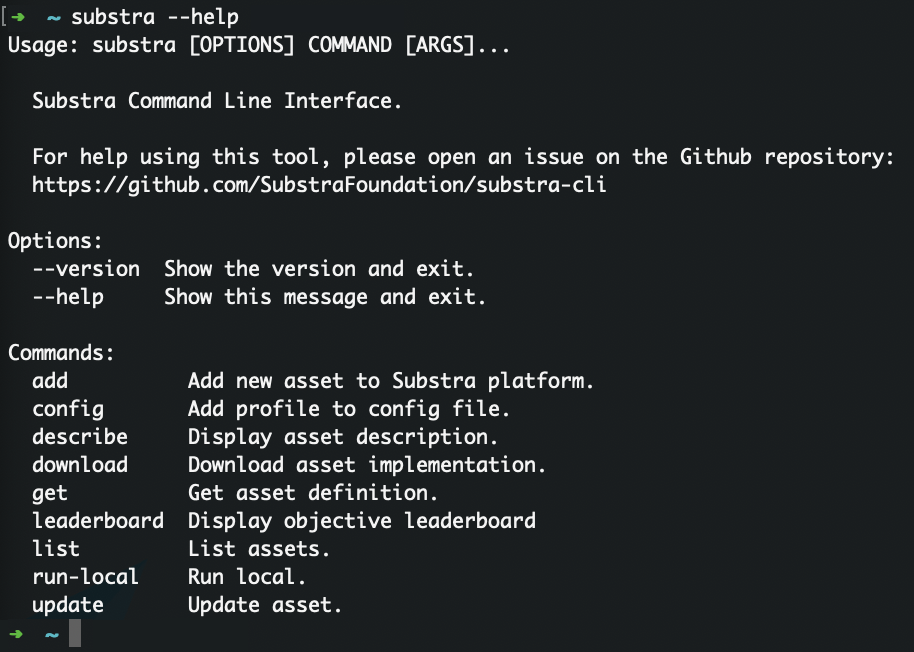} 
      		  \caption{Screenshot of the CLI help.}
     		   \label{fig: cli}
  		 \end{figure}
  		 
    \item The \textbf{Python SDK} provides the same functionalities as the CLI and offers the flexibility of a Python environment.
    As Python is a favored programming language of data scientists, it makes it easier for them to interact with Substra.
    In addition, the SDK makes it possible to integrate Substra in any Python-based application. 
\end{itemize}

\section{Architecture}\label{sec: architecture}

Substra is a distributed software to orchestrate ML computation under tight privacy constraints. As opposed to the classical client/server architecture, Substra is fundamentally decentralized.
It orchestrates the remote execution of ML tasks across several data centers.
By design, the data never leave their original servers.

Inherently, Substra provides full traceability and control of data usage.
At its core lies a decentralized and trustless consensus network which guarantees that all operations are orchestrated and written in an incorruptible ledger.
No party can modify the ledger individually. The platform exclusively relies on the ledger to dictate its behavior, thus providing strong guarantees of traceability and reproducibility.
The ledger in each node is identical.
It records the history of all past, present, and schedulded operations on the network.

The various permission regimes individually governing each asset are also stored in the ledger as smart contracts.
They have a regulating effect on the ledger in that the permission regimes filter the items added to the ledger.
Every operation must meet all the permission constraints before it is added to the ledger and then subsequently executed by the platform. Since the smart contracts are self-enforceable, i.e. they do not rely on any third party to be executed, permission constraints are met by design.

Substra can be viewed as a secure API to run ML computations on third party data. In particular, it does not include automated design and execution of coherent training strategies over distributed datasets. Substra simply receives orders from users, checks their permission, and executes them.

\subsection{Network architecture}
\begin{figure}[htbp] 
    \centering
    \includegraphics[width= 0.5\textwidth]{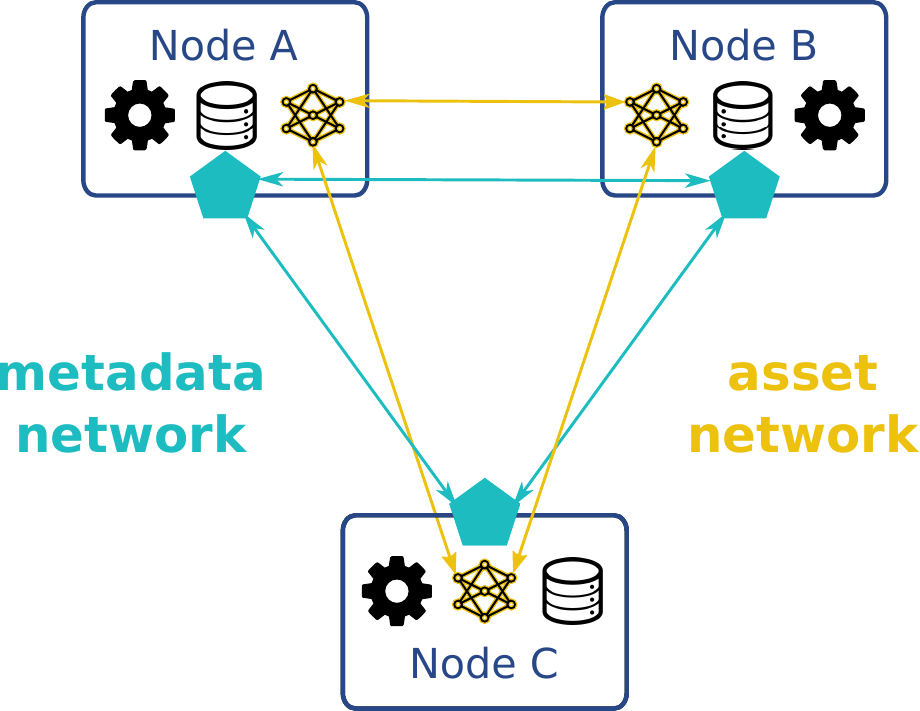}
    \caption{Global architecture of a Substra network.}
    \label{fig:global network archi}
\end{figure}

Substra is decentralized: it runs on, and connects to, a set of machines in a private network. It is made of three parts: distributed nodes, a metadata network, and an asset network as shown in figure \ref{fig:global network archi}.

\vspace{7cm}
\begin{itemize}    
    \item The \blue{\textbf{Nodes}} are mades of four components\\
        \begin{minipage}[c]{0.6\textwidth}
            \begin{itemize}
                \item \textbf{Computing resources} can access private data and common assets in order to perform containerized computations.
                The computations are engaged only if specified in the ledger.
                \item Storage of \textbf{private data} which never leave the node.
                This usually corresponds to raw sensitive data (e.g. medical data) which should remain private at all cost.
                The data are secured and used exclusively by the computing resources of the node.
                \item Storage of \textbf{common assets} such as algorithms or trained models, which can be shared between nodes under permission constraints.
                They are used exclusively by the computing resources of the node. 
                \item A shared and immutable \textbf{ledger} which stores all the operations on the platform from computing tasks to data or models registration, and the complete permission settings associated to each \textit{Asset}.
                Only non-sensitive metadata are stored in the ledger, as detailed in section \ref{ledger}.
                A library of smart contracts, called the chaincode, is used to read and write to the ledger. 
            \end{itemize}
        \end{minipage}
        \begin{minipage}[c]{0.4\textwidth}
                    \centering
                    \includegraphics[width=0.8\textwidth]{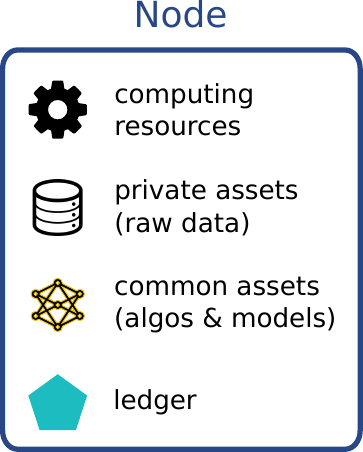}
        \end{minipage} 
    \item An  \yellow{\textbf{asset network}} for selectively sharing common assets\\
        \begin{minipage}[c]{0.6\textwidth}
            \begin{itemize}
                \item The networks makes it possible to exchange authorized models and algorithms between nodes. More generally it is a channel to communicate large files with selected partner nodes. 
                \item Asset download is restricted for use by the node computation unit and by strict permission rules. All assets have explicit permission regimes which are checked systematically before manipulating them.
                \item Private assets, e.g. sensitive raw data, are never shared on the asset network.
            \end{itemize}

        \end{minipage}
        \begin{minipage}[c]{0.4\textwidth}
                    \centering
                    \includegraphics[width=0.8\textwidth]{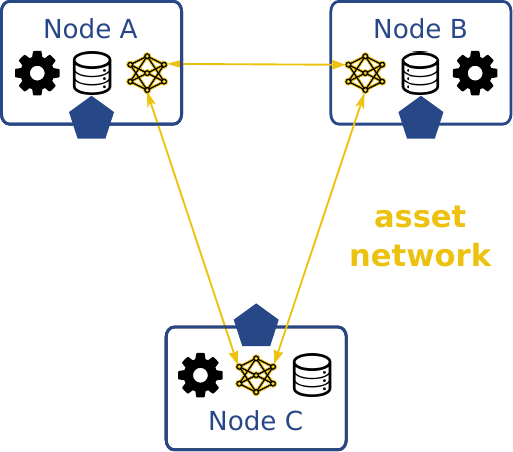}
        \end{minipage} 
    \item A \turquoise{\textbf{ledger network}} for sharing and updating the ledger \\
        \begin{minipage}[c]{0.6\textwidth}
            \begin{itemize}
                \item It is powered by a DLT framework: Hyperledger Fabric \cite{Hyperledger-ref}. The ledger is consensually built and can not be corrupted. It is operated by the chaincode mentioned above. 
                \item The ledger of each node is updated frequently and consistently in order to register new \textit{Assets}, set the \textit{Asset} permissions, and append recent/requested computation tasks to the task history. As mentioned above, only non-sensitive metadata transit within this network.
            \end{itemize}
        \end{minipage}
        \begin{minipage}[c]{0.4\textwidth}
                    \centering
                    \includegraphics[width=0.8\textwidth]{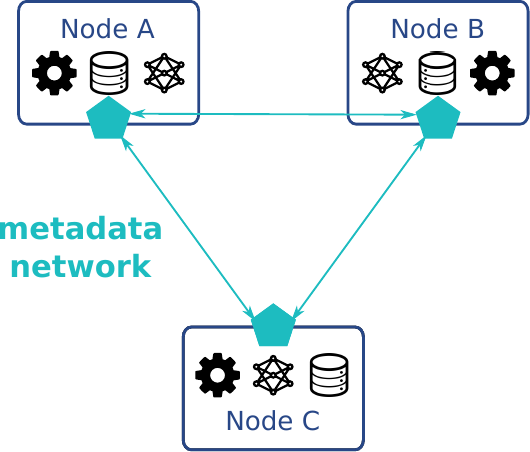}
        \end{minipage} 
\end{itemize}

\begin{figure}[H]
    \centering
    \includegraphics[width= 0.8\textwidth]{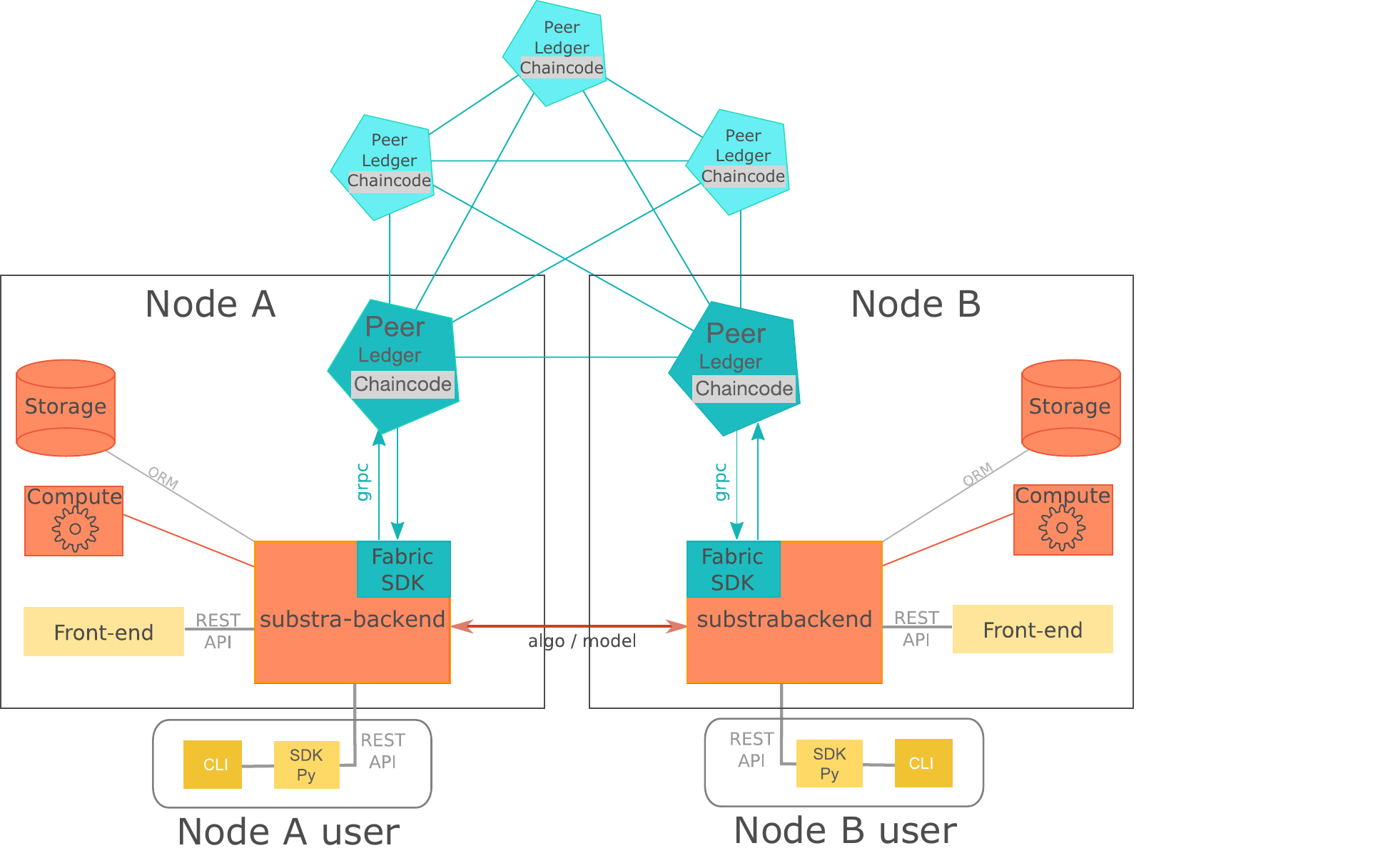}
    \caption{Overview of the architecture of Substra, illustrated with two nodes. Turquoise arrows indicate exchanges of non-sensitive metadata between the backend of a node and the ledger network. The red arrow indicate the exchange of an algorithm or model through the asset network.}
    \label{fig:detailed archi}
\end{figure}

Figure \ref{fig:detailed archi} details the architecture of Substra, and interactions between its different components.

\subsection{Workflow}
The workflow describes the typical sequence of steps that are performed by the platform in order to address user requests properly.
It is out of the scope of this document to detail all of them, only the general and principal patterns are presented here.

\subsubsection{Orchestration by the ledger of computations}

The workflow of a Substra network is driven by two coupled and asynchronous circuits: ledger and computation operations.
\begin{itemize}
    \item Ledger operations are performed through the DLT framework and are triggered by user inputs. They mainly consist in registering assets and specifying (possibly a sequence of) computations to be performed independently.
    \item Computation operations are automated and performed locally in the node with private and common assets.
    Computations are only triggered and authorized when defined in the ledger.
    Computation results are logged in the ledger.
\end{itemize}
Thus, if a node is randomly removed from the network then the other nodes can continue operating normally, except that they lose their ability to process the private data of the leaving node.

\subsubsection{Example of workflow}
\begin{figure}[H]
    \centering
    \includegraphics[width= 0.7\textwidth]{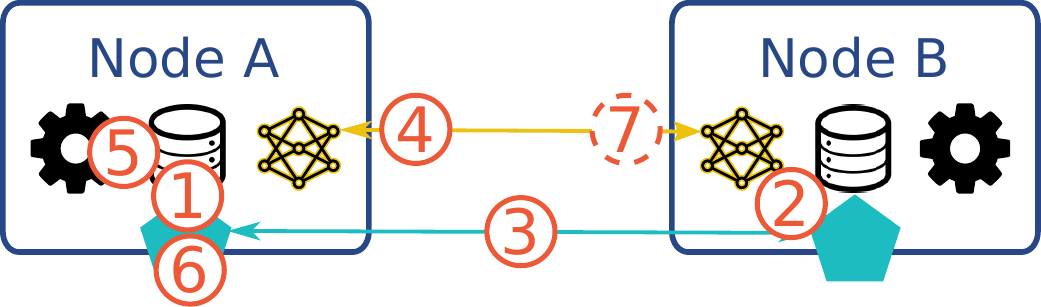}
    \caption{Substra's workflow between two example nodes. See description of steps in text.}
    \label{fig:workflow}
\end{figure}

To help understand the detailed workflow, a simplistic yet representative example between two nodes is detailed below.
This corresponds to user A (owner of node A) being a data controller (e.g. a hospital) and user B (owner of node B) being an algorithm developer (e.g. an AI company).
Both parties explicitly agree through the permission settings that user B can train her algorithms on user A's private data.
Figure \ref{fig:workflow} illustrates the successive steps in the processing of user's B request.
\begin{enumerate}
    \item User of node A registers private data to the platform. Associated non-sensitive metadata are submitted to the ledger (see details in section \ref{ledger}). The ledger update is automatically broadcasted and accepted by the other node.
    \item User of node B registers a common/shared algorithm to the platform. Associated non-sensitive metadata are submitted to the ledger. The ledger update is automatically broadcasted and accepted by the other node.
    \item User of node B requests a computation of his algorithm on user A's private data. A new computation task is specified in the ledger. It is then broadcasted and accepted by the other node only if it meets the permission settings of node A's private data, which are found in the ledger.
    \item Node A observes that there is a new task to be processed in the ledger and automatically downloads the common algorithm from node B (which authenticates node A against the ledger) and stores it.
    \item Node A securely performs the computation and applies user B's algorithm on its own private data.
    \item Non sensitive metadata summarizing the computation execution and resulting performance are written by Node A in the ledger which synchronizes with other nodes immediately. 
    \item (Optional) Outputs of the computations, such as trained models (but never private data), can be sent back to node B depending on chosen permissions.
\end{enumerate}
This simple workflow can be chained and composed to perform arbitrarily complicated computations at the scale of the network, as detailed in section \ref{compute plans}.  

The ledger gathers all user inputs and the computation units listen to ledger updates in order to trigger and perform computations on their private data.

\subsection{Information stored in the ledger} \label{ledger}

As mentioned previously, the ledger stores only non-sensitive metadata, required for the orchestration and for the traceability of the training of machine learning models on distributed data.

Table \ref{tab:ledger} summarizes elements stored in the ledger.

\begin{table}[H] \label{tab:ledger}
\begin{tabular}{|p{3cm}|p{12cm}|}
\hline
\textbf{Object} & \textbf{Attributes stored in the ledger} \\ \hline
\textit{Objective} &
- Name  of the \textit{Objective}
\newline - Storage address and hash of its description
\newline - Name, storage address and hash of its metrics
\newline - Owner (node who defined the objective)
\newline - Test datasets (list of data keys for the test split, and their associated dataset)
\newline - Permissions
\\ \hline

\textit{Dataset} &
- Name of the \textit{Dataset}
\newline - Storage address and hash of its data opener
type of data in the dataset (tabular, image, ...)
\newline - Storage address and hash of its description file
\newline - Owner
\newline - Associated \textit{Objective} key
\newline - Permissions
\\ \hline 

Data&
- Hash of data stored in local storage
\newline - List of keys of associated \textit{Datasets}
\newline - A boolean indicating if data is dedicated to testing
\\ \hline

\textit{Algorithm} &
- Name of the \textit{Algorithm}
\newline - Storage address and hash of the algorithm files
\newline - Owner
\newline - Associated \textit{Objective} key
\newline - Storage address and hash of the description of the algorithm \newline - Permissions
\\ \hline

\textit{Traintuple}&
- Associated \textit{Objective} key (for its metrics)
\newline - Associated \textit{Algorithm} key
\newline  - List of input models (list of traintuple keys, hashes, addresses)
\newline - Output \textit{Model} (hash, address)
\newline - List of training data and the node where they are stored \newline - Status of the task: waiting, todo, doing, done, failed
\newline - Log
\newline - Optional arguments necessary for complex ML orchestration (a rank and a tag)
\newline - Permissions
\newline - Creator (node who defined the traintuple)
\\ \hline

\textit{Testtuple}&
- Associated \textit{Objective} key (for the metrics)
\newline - Associated \textit{Algorithm} key
\newline - Model to evaluate (hash, address)
\newline - List of testing data and the node where they are stored \newline - Status of the task: waiting, todo, done, failed
\newline - Log
\newline - Optional arguments necessary for complex ML orchestration (e.g. a tag regrouping several ML tasks)
\newline - Permissions
\newline - Creator (node who defined the testtuple)
\\ \hline
\end{tabular}
\end{table}

%    \item A \textit{Dataset} aggregates numerous data points under a single format.
%    It includes a single \textit{Opener} script which imports and opens the file using libraries specific to the data type.
%    \item An \textit{Algorithm} is a script which specifies the method to train a \textit{Model} on a \textit{Dataset}. In particular it specifies the model type and architecture, the loss function, the optimizer, hyperparameters and identifies the parameters that are tuned during training.
%    \item A \textit{Model} is a large file containing the parameters of a trained model.
%    In the case of neural networks, it gathers the weights of the connections.

\section{ML orchestration} \label{compute plans}
To launch Substra on large amounts of distributed data, users must create sequences of tasks which are executed by Substra.
These are called \textit{compute plans}.
They prescribe unambiguously the organization and ordering of computations for training algorithms on datasets.
They also make it possible to evaluate a model against several test datasets.
Substra does not involve automated generation of \textit{Compute plans}.
It simply executes the user's orders in the form of \textit{Compute plans}.

\subsection{Chaining training and averaging}
The first building block of \textit{compute plans} is the \textbf{training step}, specified by a \textit{traintuple}.
It defines a unitary training task which updates a \textit{Model} as the result of the training of an \textit{Algorithm} on a given \textit{Dataset}.
In the classical setup \textit{traintuples} take a single model as input and provide another model as output.

The second building block of \textit{compute plans} is the \textbf{averaging step}.
It takes several \textit{Models} as inputs and outputs a single "averaged" \textit{Model}, either by averaging the prediction of the input models \cite{claeskens2008model}, or by averaging the weights of the inputs models \cite{mcmahan2016communication}.
Depending on the type of model used and  the particular FL strategy, more complex operations might be needed. 

\begin{figure}[htbp]
    \centering
    \includegraphics[width= 0.6\textwidth]{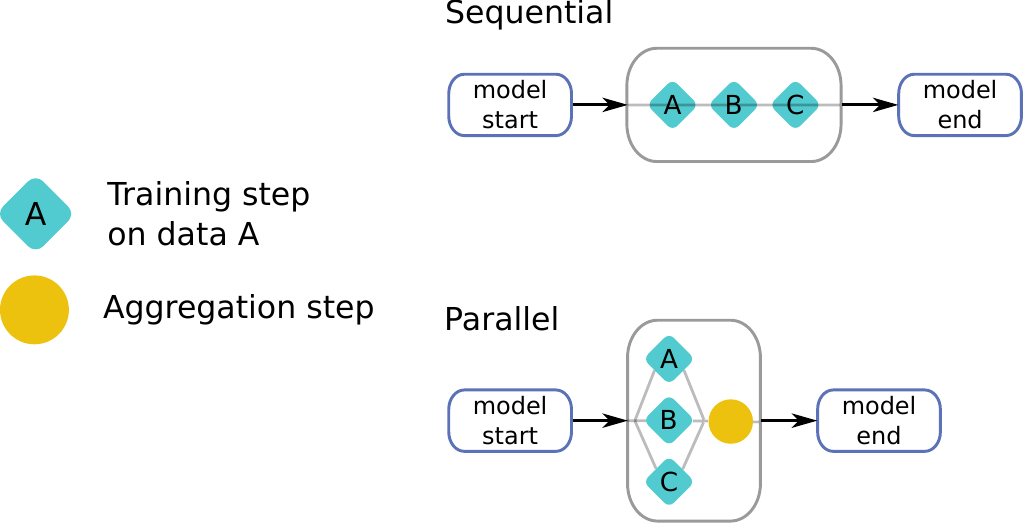}
    \caption{(left) Elements of \textit{compute plans} (see text for details). (right) Basic \textit{compute plan} samples: (i) sequential training consists in a sequence of training on different datasets, (ii) parallel learning is made by training several models independently but with the same initial model, then an aggregating step merge the several models in a single output.}
    \label{compute plan 1}
\end{figure}

A \textit{compute plan} is a set of training and averaging steps, whose in and out models are chained in a specific and possibly complicated pattern.
Basically, chaining can be done in a sequential or parallel way as illustrated in figure \ref{compute plan 1}.

\textit{Sequential compute plans} only use training steps; there is no averaging involved. In the simplest case, the \textit{Algorithm} is fixed and a \textit{Model} is successively trained on multiple remote \textit{Datasets}.
Note that changing the order of the datasets is likely to change the end model.

\textit{Parallel compute plans} correspond to a sucession of training and averaging steps. First, several models are trained from the same initial model using different training datasets. Second, the models are aggregated in a single output model.

\begin{figure}[htbp]
    \centering
    \includegraphics[width= 0.6\textwidth]{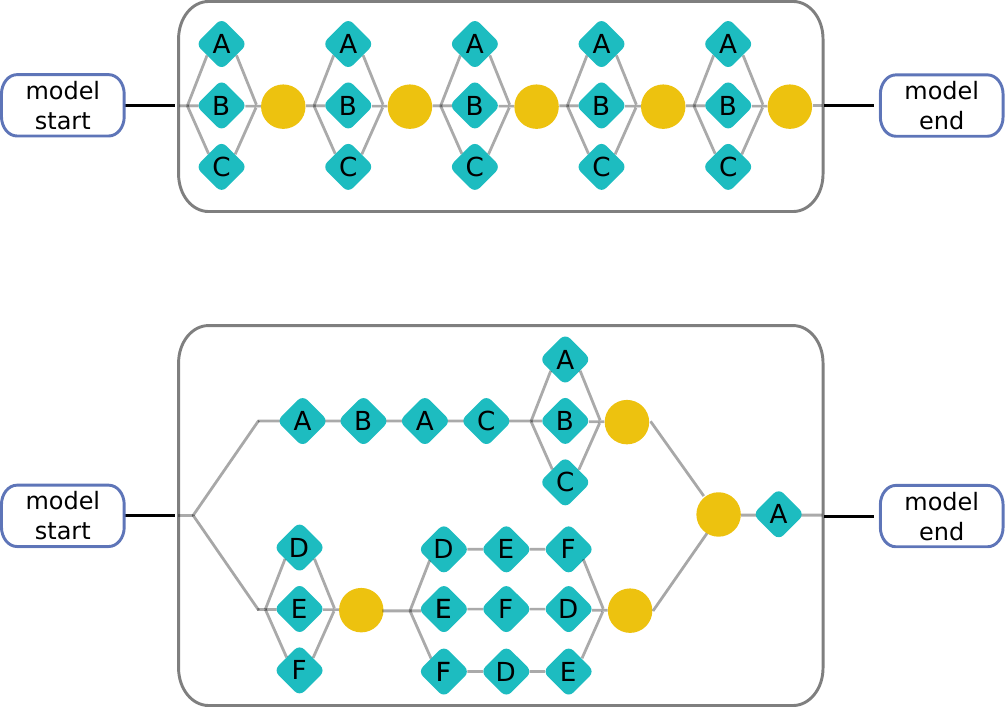}
    \caption{Sample \textit{compute plans}. (top) Standard \textit{compute plan} with regular succession of parralel and averaging steps. (bottom) Excessively complicated sample \textit{compute plan} to illustrate the compostion of basic patterns.}
    \label{compute plan 2}
\end{figure}

These two basic patterns can be chained and composed to create arbitrarily complex \textit{compute plans} as illustrated in figure \ref{compute plan 2}.
The first \textit{compute plan} shown in figure \ref{compute plan 2} corresponds to the standard pattern in federated learning \cite{mcmahan2016communication}.

\subsection{Evaluation}
Evaluation is an important part of Substra. It is split in validation datasets for hyper-parameter tuning and test datasets for evaluation on new data.
\begin{itemize}
    \item \textbf{Validation} is made flexible and entirely parametrable by users who can perform any kind of cross validation scheme on any subpart of the train datasets.
    \item \textbf{Test} datasets however are sanctuarized and can never be used for training. Test performance evaluation is constrained to a rigorous methodology.
\end{itemize}

\begin{figure}[htbp]
    \centering
    \includegraphics[width= 0.6\textwidth]{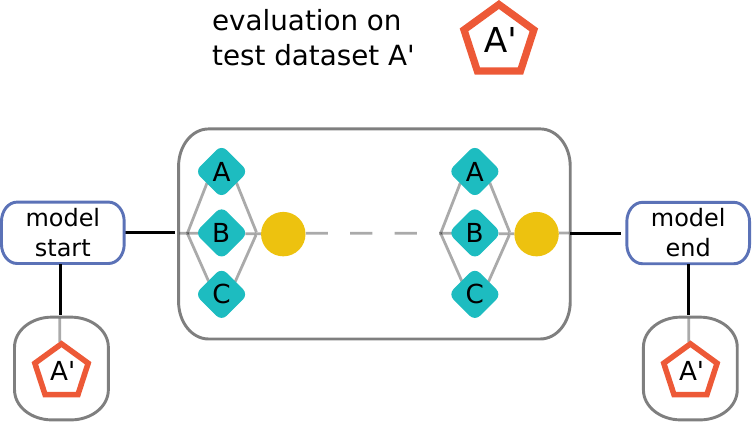}
    \caption{Sample \textit{compute plan} involving evaluation steps. Here the owner of data A and A' is training an algorithm on other remote datasets. Although it involves other data, the evaluation is only made on A'.}
    \label{compute plan 2 eval}
\end{figure}

Each node in the network may define a subpart of its data as an immutable test dataset.
The evaluation against these datasets is defined in the form of an \textit{Objective} and can be requested by any user with the appropriate processing rights.
Figure \ref{compute plan 2 eval} illustrates a typical evaluation pattern of the performance of a model before and after training.
Notably, the evaluation in Substra can be done on each user dataset.
Thus, there may be different performance levels of a single model depending on the node/user evaluating it.

Cross-validation is a standard methodology for evaluating model performance. For a k-fold cross-validation, k pairs of training and test datasets are derived from the original dataset, and the performance is computed as the average of the performance on the k test datasets.
In Substra cross-validation is formalized as a specific \textit{Objective} with a training dataset, but without test dataset.

\subsection{Model composition}
Large ML models, in particular deep learning models, are often the composition of several sub models which can be trained independently.
Substra supports models defined as combinations of others.
Of course, the complexity of the resulting algorithm has to be handled by the user; but Substra has been specifically designed to make it easy to handle training tasks over combined models.

\begin{figure}[htbp]
    \centering
    \includegraphics[width= 0.9\textwidth]{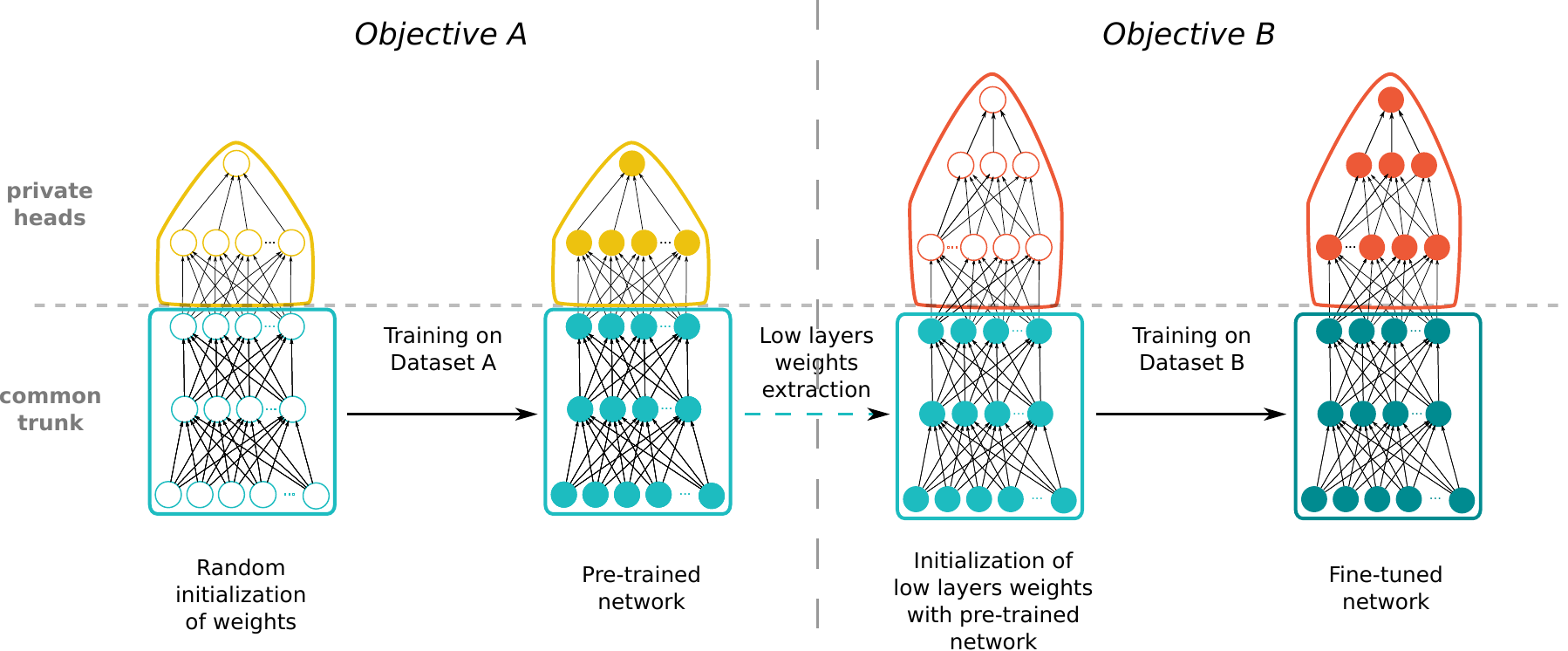}
    \caption{Classical multi-task learning approach called warm restart.
    First, a neural network is trained on \textit{Dataset} A and evaluated against \textit{Objective} A.
    The initialization of the weights is usually random. In order to transfer the knowledge from the pre-trained \textit{Model} to another task, one can extract its lower layers and consider them as the initial weights of a second network built for \textit{Objective} B, which is to be fine-tuned on \textit{Dataset} B.
    In this case the inputs are considered the same across \textit{Objectives}.}
    \label{fig: multitask simple}
\end{figure}

Transfer learning on neural networks often implies modifying an existing network by adding or removing some layers and fine tuning some connections in the network.
For instance, when considering a task of classification over images, it is common to download the weights of a pre-trained neural network designed for discriminating other classes (such as ResNet \cite{he2016deep}) and recycle the lower layers of the deep network as shown in figure \ref{fig: multitask simple}.
This corresponds to what is called a warm start. It makes it possible to reuse the weights trained for a certain \textit{Objective} for another one.
This is the canonical example of sequential orchestration with model composition. In fact, figure \ref{fig: multitask simple} corresponds to a simple sequential \textit{compute plan}.

This means several users can decide to share exclusively the lower layers of a network.
This would correspond to changing the \textit{process permission} of the sub-network.
This common subpart of the \textit{Model} is called a \textit{Trunk}.
Each node keeps the upper layers of its \textit{Model} private.
They constitute the private \textit{Heads}.
Figure \ref{fig: multitask simple} illustrates a situation where two partners want to train together a common trunk model, but do not want to share their data and not even the definition of their own \textit{Objective}.
It is a situation with increased privacy where partners do not know what the other is computing.
Nonetheless, in using the backpropagation training algorithm not only through head layers, but also through trunk layers for each partner, the trunk model can benefit from the information of all partners without revealing private information.

\begin{figure}[htbp]
    \centering
    \includegraphics[width= 1.\textwidth]{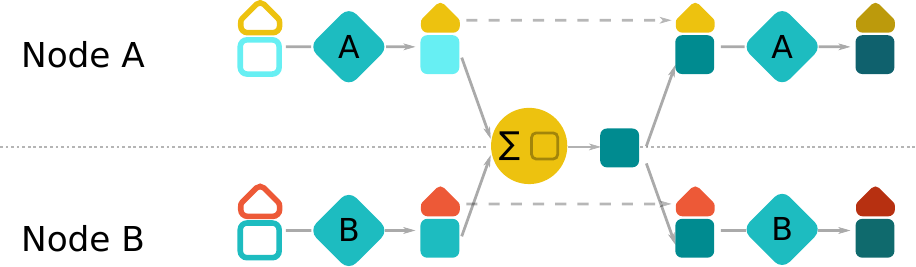}
    \caption{Federated multitask learning. The trunk model is trained collectively but the private head models are never shared between partners.}
    \label{compute plan 1_bis}
\end{figure}

Substra is also designed to tackle parallel orchestration combined with model composition.
A standard compute plan associated with such a combined model is illustrated in figure \ref{compute plan 1_bis}.

Note that a quasi identical approach can be applied for domain adaptation \cite{pan2009survey}, which corresponds to multiple users having an identical goal (and thus a single \textit{Objective}), but with data from different sources and with slightly different format or distribution.
In this case, the modularity of the network occurs at the lower layers of the neural network. 

\section{Risk analysis} \label{sec: risk analysis}
This section proposes a high-level risk analysis focused on the Confidentiality, Integrity, and Availability \cite{CIA} of the different \textit{Assets} of Substra.
An exhaustive risk analysis is out of the scope of this whitepaper, and we focus on the risks which we have estimated as the most relevant.
In particular, we focus on the risks due to an attacker being part of the network, i.e. the attacker is assumed to control entirely its own node and is willing to attack the \textit{Assets} of others.
For simplicity, we do not address the risks due to attackers from within the node or from outside the network.
These kinds of attacks are relevant and should be mitigated, but are not specific to Substra, thus they are not addressed here.

An important feature of Substra is trustless traceability.
Any risk is at least partly mitigated by transparency mechanisms.
It is not possible to launch operations which are not stored in the ledger with Substra.
Thus, any attack triggering specific computations will be recorded in the ledger for possible future inspection.

\subsection{Confidentiality}
Substra is deliberately developed to provide high confidentiality of \textit{Datasets}.
In many situations confidentiality is synonymous to privacy, which is terminology specific to personal data.
At the core of Substra design is the principle of never moving \textit{Datasets} between nodes.
Only the \textit{Algorithms} and \textit{Models} are exchanged between nodes.
Thus, compared to a classical centralized architecture, using Substra decreases the confidentiality risk on \textit{Datasets} at the expense of an increased confidentiality risk on \textit{Algorithms} and \textit{Models}.
Substra implies transfer of risk from \textit{Dataset} to \textit{Algorithms} and \textit{Models}.

The confidentiality attacks can be organized in 3 groups: \textit{Dataset} theft, \textit{Algorithm} / \textit{Model} theft and metadata leak. They are identified in the table below:
\begin{table}[H]
    \begin{tabular}{|p{1.5cm}|p{3cm}|p{5cm}|p{4cm}|}
    \hline
    \textbf{Risk} & \textbf{Risk description} & \textbf{Built-in Substra mitigation} & \textbf{Additional mitigations needed}                                                                           \\ \hline
    \multirow{2}{*}{Data theft}  & Attacker downloads raw data. & Data remain within each node infrastructure. &                                                                                                         \\ \cline{2-4}
                                                                           & Attacker designs malevolent \textit{Algorithm} to extract data (possibly in the \textit{Model} weights).  
                                                                           & \textit{Model}'s permission regime can forbid model download.

                                                                           Securing the compute worker (e.g. no access to network during computations).
& \textit{Algorithm} certification by third party.

\\  \hline
Data inference
& Attacker infer properties about the data from the trained models.
& 
Model access can be restricted.
& Certification for designing non-identifying models.

Contractualisation between partners.      \\  \hline
Algorithm / Model theft
& Attacker steals the algorithm or model during a training step.
& Securing the compute worker (e.g. model never stored on hard drive).

Permission regime selects who can process model.
& Trusted Execution Environments (TEE).

Contractualisation between partners.      \\  \hline
Metadata leak & The common ledger containing the platform metadata is leaked outside the network.
& Metadata are anonymous, the ledger only contains hashes of assets.
&
\\ 
                                     \hline
    \end{tabular}
    \end{table}

Overall, the risk of a \textit{Dataset} confidentiality breach is largely mitigated by the Substra computation architecture.
The main residual risk lies in malevolent \textit{Algorithms} which could leak \textit{Datasets} out of the node, for instance, in writing the data themselves in the model weights.
It can be strongly mitigated by requiring the \textit{Models} permission regimes to exclude download access to the corresponding \textit{Algorithm} designer.
Independent audit and certification of the \textit{Algorithm} before deployment could be a pragmatic alternative.
A more limited residual risk consists in infering properties about the individual data from the trained model itself.
This is actually unlikely if the training procedure is well designed and the model diffusion limited.

The residual risk on \textit{Algorithms} and \textit{Models} confidentiality is not negligible.
On its own, Substra can only make it difficult for a node to steal a \textit{Model} which is being trained locally.
Since the program runs on the machine of the user it cannot be theoretically bulletproof.
A future idea for a solution would be to rely on the growth of Trusted Execution Environments \cite{sabt2015trusted, lee2019keystone}, but the lack of required GPU support for current TEE make this possibility speculative. 
In practice, it is necessary to consider contractual interactions between the network partners to cover the residual risk on \textit{Algorithms} and \textit{Models}.
Today, Substra is not likely to be deployed in an open network where everyone could participate as in a public blockchain.

Confidentiality of \textit{Datasets} goes beyond the simple access to the raw data; privacy of individuals who are part of the \textit{Dataset} is to be guaranteed.
The design of Substra was strongly influenced by the GDPR and takes the question of privacy as a first principle.
More precisely, an important risk to address is the leaking of high level personal information without accessing the \textit{Datasets} themselves.

There are several guidelines or good practices that we strongly recommend to adopt when deploying Substra over personal data. Their details are out of the scope of the whitepaper.
\begin{itemize}
    \item Always \textbf{pseudonymize} data before registering them as \textit{Datasets} in Substra.
    Anonymization procedures \cite{g29anon-ref} are even better, but not always relevant for certain use cases (e.g. Healthcare where re-identifying patients is key to provide personalized care).
    \item Design \textbf{non-identifying} \textit{Models}, so that they can be exchanged without exposing personal information. \textit{Models} are statistical objects which do not need to contain identifying information.
    Actually, it is often the case that models that leak information about individual data points are ill-designed for this setting; for instance in the classical problem of overfitting (\cite{yeom2018privacy}).
    Requiring training \textit{Algorithms} to be properly regularized is paramount. Similarly, an efficient mitigation consists in restricting training steps to be performed on large groups of individuals so as to dilute the personal information in the aggregate.
    Trusted audits and certification of \textit{Algorithms} are necessary for the use cases requiring the highest degree of privacy.
\end{itemize}
In other words, Substra does not solve the problem of privacy on its own; a number of privacy-enhancing protocols must be enforced to guarantee the highest level of privacy.

\subsection{Integrity}
Substra provides high integrity guarantees, mainly because of its architecture rooted in an unfalsifiable distributed ledger logging all ML operations.
All \textit{Assets} are registered and referenced through a unique identifier.
The identifier is in fact the hash of the \textit{Asset}, which makes it possible to guarantee the \textit{Assets} have not changed when reproducing a sequential training procedure.
Integrity of \textit{Assets} can therefore be checked at all times.

Integrity applies also on the results, i.e. predictions on new data.
It is crucial to make sure that the predictions cannot be biased by an attacker.
Beyond standard attacks on a node which are not covered here, there is a particular kind of attack fundamentally linked to the specifics of Substra: participating in a collaborative training (with \textit{Datasets} or \textit{Algorithms}) in order to bias the prediction of the output \textit{Model}.
A possible mitigation of this risk is to test the predictive performance on specific test \textit{Objectives} equipped with controlled and independent test \textit{Datasets}.
It is indeed likely that methods devoted to bias prediction lead to poorer performance on test datasets.
There is a clear security incentive to consider and favor the best \textit{Model} for each \textit{Objective}. However, a residual risk lies in the fact that large predictive models could contain stolen data while keeping good performance.

\begin{table}[H]
    \begin{tabular}{|p{1.5cm}|p{3cm}|p{5cm}|p{4cm}|}
    \hline
    \textbf{Risk} & \textbf{Risk description} & \textbf{Built-in Substra mitigation} & \textbf{Additional mitigations needed}
\\ \hline

Results integrity 
& Attacker changes the performance metadata.
& DLT prevents from changing reports made by others.
& 
                  \\ \hline
Results and \textit{Assets} integrity
& Attacker modifies a model to alter prediction quality (for instance by training on biased datasets).
& Full traceability of operations and storing of intermediate models to be able to revert a bad training step.
& Independent test sets to validate performance evolution                  \\ \hline
    \end{tabular}
    \end{table}

\subsection{Availability}
Availability attacks are not critical to Substra functioning.
Being a distributed, asynchronous network, Substra has a low-latency.
This is not a significant problem since Substra is focused on training \textit{Models} which does not involve immediate user interaction.
It is also resilient to attacks on single nodes due to its decentralized architecture: an availability attack on a partner only blocks the \textit{Assets} exclusively owned by this partner.

To decrease the risk of overloading a node, the ledger or the network with too many training or prediction tasks, permission regimes can be set up to create a white list of users which can request computations.
The permissions are applied a priori to prevent illegitimate training tasks to be added in the ledger and synchronized over the network.
\begin{table}[H]
    \begin{tabular}{|p{1.5cm}|p{3cm}|p{5cm}|p{4cm}|}
    \hline
    \textbf{Risk} & \textbf{Risk description} & \textbf{Built-in Substra mitigation} & \textbf{Additional mitigations needed}
    \\ \hline
    Service availability
    & Attacker overloads a node with training or prediction requests.
    & Other nodes still function.
     
    Permission regime authorizes only specific nodes to launch computations on one's node.
    & 
    \\ \hline
    \end{tabular}
    \end{table}

\section{Perspectives}
Substra's ambition is to become the standard framework for performing collaborative ML over distributed datasets.
It is designed to be modular and open.
There are countless ways to improve the framework or integrate it into existing software solutions.

\subsection*{Integration with similar projects}
There are several technologies similar to Substra which are emerging today; we believe the future of Substra is conditioned by its ability to interface nicely with other software in particular in the open source community. Collaborative initiatives between projects are likely to drive the rise of a "responsible and trustworthy data science" ecosystem that we pledge for.

To name only a few similar projects: 
\begin{itemize}
    \item OpenMined \cite{openmined} is building PyTorch libraries for privacy-preserving deep learning with deep training content \cite{udacity-course}.
    \item Dropout Labs \cite{dropoutlab} is building TensorFlow libraries for ML on encrypted data.
    \item Google is designing a library/SDK for TensorFlow to add federated learning features for smart phones \cite{google-federated}.
    \item Ocean Protocol \cite{oceanprotocol} and Oasis Labs \cite{oasislab} are building decentralized networks enabling the development of privacy-preserving data-based applications.
\end{itemize}

\subsection*{Secure aggregation of model updates}
In parallel compute plans, there is a central aggregator which takes many model updates from different nodes as inputs and outputs a single averaged model.
This is a typical pattern in Federated Learning \cite{mcmahan2016communication}.
It is associated with a significant risk at the central point since the aggregator has access to all model updates from the nodes, from which it could infer sensitive information about the nodes' data.
A classical approach to mitigate this risk is to use a mechanism of secure aggretion where the model updates are individually obfuscated by the aggregator can still output an accurate average \cite{bonawitz2017practical}.
This is an interesting perspective for Substra for some use-cases.

\subsection*{Non assignability of metadata}
Substra could benefit from strong anonymity features regarding who owns which \textit{Asset}.
For now, assigning the ownership of an \textit{Asset} to a node is accessible to competent and technical attackers.
However, the DLT-based architecture of Substra gives us reason to assume that providing anonymity is technically possible.
Substra is still a consortium based technology; thus it will only provide anonymity among the members of an explicit consortium. But "hiding among the trees" can be relevant to numerous use cases. 

\subsection*{Ownerless \textit{Assets}}
Substra could open the possibility of having ownerless \textit{Assets}.
By combining a Multi Party Computation approach (such as Shamir secret sharing) with the ledger of Substra, one could imagine encrypting some \textit{Assets} and split the decryption key among nodes.
Later when a \textit{traintuple} involves this \textit{Asset}, a  node could gather all the key parts from other nodes to decypher the \textit{Asset} which could be erased from this node after computation.
This feature could open brand new use cases where extra privacy or decentralized control of \textit{Assets} is required.
This raises difficult technical and security challenges, for instance, to keep the availibility of service sufficiently high for practical applications.

\subsection*{Non-identifying \textit{Models}}
As shown in the risk analysis in section \ref{sec: risk analysis}, the usage of Substra needs to be considered together with the definition and adoption of sound guidelines for \textit{Algorithm} design in order to reach the highest level of privacy preservation.
The guidelines should guarantee that  \textit{Models} are non-identifying. This means that the precise information about data samples in the train \textit{Datasets} cannot be retrieved from a \textit{Model}.
For instance, these guidelines will surely address the problem of overfitting for \textit{Model} design.
This problem is not only bad for generalization of \textit{Models}, but it also leaks much more information about the overfitted \textit{Dataset}.
Importantly, these guidelines will have to be promoted and checked by trustful regulatory entities.

\subsection*{Partner ecosystem}
On an ecosystemic aspect, the growth of Substra into a widespread, production-grade network is likely to involve different roles among the partners within a Substra network. 
Inevitably, the growth of Substra will be bounded by the capacity of the current actors of our socio-economic ecosystem to structure themselves to use the technology to their specific advantage.
A first category of actors gathers \textit{Asset} controllers.
Some of them will specialize in \textit{Datasets} collections and management (e.g. Hospitals), whereas some others will specialize in \textit{Algorithm} creation and management (e.g. AI startups).
Both types of actors will inevitably have to deal with \textit{Model} management and permission setting. A second category of actors gathers regulators and evaluators of the \textit{Models} performance.
They will deal with the design and management of \textit{Objectives} and in particular the associated test \textit{Datasets}.
Note that it is crucial to have independent actors providing standardized benchmarks of \textit{Models}.

\subsection*{Token-based economic ecosystem}
Finally, an interesting perspective would be to leverage the built-in traceability within Substra in order to create a token-based economic system for collaborative Machine Learning. This would involve attributing a value to the Substra \textit{Asset} which could be aligned with the performance improvement brought by each \textit{Asset} on predefined \textit{Objectives}. For instance, one could compute a contribution score for each \textit{Datasets} by evaluating the percentage of improvement observed after training on it. Although it is beyond the scope of the current version of Substra, this economic system could be a real driver for the growth of Substra and collaborative, privacy-preserving Machine Learning.

\section*{Acknowledgement}
This work is funded by Owkin and Bpifrance as part of the "Healthchain" project, which resulted from the "Digital Investments Program for the major challenges of the future" RFP. 

We are grateful to the open source community for all the great libraries at the basis of our work.
Standing on the shoulders of giants as they say.

Special thanks to Eric Boniface, Kelvin Moutet, Guillaume Cisco, Eric Tramel, Samuel Lesuffleur, Cl\'ement Gautier, Romain Bey, Romain Goussault, Inal Djafar, Cl\'ement Mayer, and Talia Litteras for helping improve the manuscript.

We thank the developers of Substra, the members and staff of Substra Foundation and the employees of Owkin for all the useful discussions, experiments, and contributions that led to this document and the associated source code.

\bibliographystyle{plain}
\bibliography{substra_whitepaper_arxiv}
\end{document}